\documentclass[aps,prl,twocolumn,superscriptaddress, showpacs]{revtex4}
\usepackage{amsmath}
\usepackage{graphicx}
\usepackage{fancyhdr}
\usepackage{color}

\begin{document}


\title{Rapid Domain Wall Motion in Permalloy Nanowires Excited by Spin-Polarized
Current Applied Perpendicular to the Nanowire}


\author{C. T. Boone}
\affiliation{Department of Physics and Astronomy, University of California, Irvine,
CA 92697}
\author{J. A. Katine}
\affiliation{Hitachi Global Storage Technologies, San Jose, CA 95135}
\author{M. Carey}
\affiliation{Hitachi Global Storage Technologies, San Jose, CA 95135}
\author{J. R. Childress}
\affiliation{Hitachi Global Storage Technologies, San Jose, CA 95135}
\author{X. Cheng}
\affiliation{Department of Physics and Astronomy, University of California, Irvine,
CA 92697}
\author{I. N. Krivorotov}
\affiliation{Department of Physics and Astronomy, University of California, Irvine,
CA 92697}
\begin{abstract}
We study domain wall (DW) dynamics in permalloy nanowires excited by alternating
spin-polarized current applied perpendicular to the nanowire. Spin torque
ferromagnetic resonance measurements reveal that DW oscillations at a pinning site
in the nanowire can be excited with velocities as high as 800 m/s at current
densities below 10$^7$ A/cm$^2$.
\end{abstract}

\date{\today}
\maketitle

Domains in ferromagnetic nanostructures are candidates for information storage
\cite{parkin2, hayashi3} and processing \cite{allwood}. To be technologically
useful, magnetic domains must be easily manipulated.  One way this can be
accomplished is by moving the magnetic domain wall (DW) separating two oppositely
magnetized domains. Since manipulation of individual DWs in nanostructures with
magnetic field  \cite{schryer, glathe} is technically challenging, DW motion induced
by spin torque (ST) from spin-polarized current has emerged as a promising
alternative \cite{berger}. The effects of ST on DW are most readily studied in the
ferromagnetic nanowire geometry with electric current applied along the nanowire
\cite{groll, yamag, beach, saitoh, klaui, thomas, ravel, thia2, tatara1}. However,
spin torque exerted on a DW in this geometry is small because the angle between the
current polarization direction and magnetization is small everywhere in the DW. As a
result, high current density ($>$10$^8$ A/cm$^2$) is required to achieve DW
velocities in the technologically useful range of $\sim$10$^2$ m/s \cite{hayashi2}.

High DW velocities (180 m/s) at relatively small current densities
($\leq5\times$10$^7$ A/cm$^2$) have been observed in spin valves with current
flowing in the plane of the magnetic layers \cite{pizzini}. This high efficiency of
the current-induced DW motion \cite{groll} was attributed to transfer of angular
momentum from the fixed layer to the DW in the free layer of the spin valve mediated
by the component of spin current perpendicular to the layers. Khvalkovskiy
\textit{et al.} \cite{khval} also numerically studied DW motion in spin valve
nanowires with current flowing perpendicular to the spin valve layers (CPP
geometry). This work predicted that DW velocities of $\sim$10$^2$ m/s can be
achieved at current densities of $\sim$10$^7$ A/cm$^2$ in the CPP geometry
\cite{rebei}. 

In this Letter we make measurements of CPP DW dynamics in the permalloy
(Py$\equiv$Ni$_{84}$Fe$_{16}$) layer of Co$_{50}$Fe$_{50}$/ Cu/ Py spin valves
patterned into 5 $\mu$m long and 90 nm wide nanowires. The nanowires are defined on
metallic Cu/Ta/Ru films serving as bottom leads, and 500 nm wide Ta/Au top leads are
used to apply current perpendicular to the spin valve layers as shown in Fig. 1. The
device in Fig. 1 is made in a multi-step nanofabrication process starting from a
Ta(5)/ Cu(30)/ Ta(3)/ Cu(30)/ Ta(5)/ Ru(10)/ Cu(3)/ Co$_{50}$Fe$_{50}$(7)/ Cu(5)/
Py(3)/ Cu(5)/ Ru(2.5)/ Ta(2.5) multilayer (thicknesses in nanometers). The
multilayer is deposited onto a thermally oxidized silicon substrate by magnetron
sputtering and annealed at 225 $^{\circ}$C for 2 hours. The Py layer magnetization
measured with vibrating sample magnetometry ($M_s$= 430 emu/cm$^3$) is reduced
compared to the bulk value due to magnetically dead layers at the Cu/Py interface
and interdiffusion of Cu and Py \cite{hecker}. 

\begin{figure}
\includegraphics[width=\columnwidth]{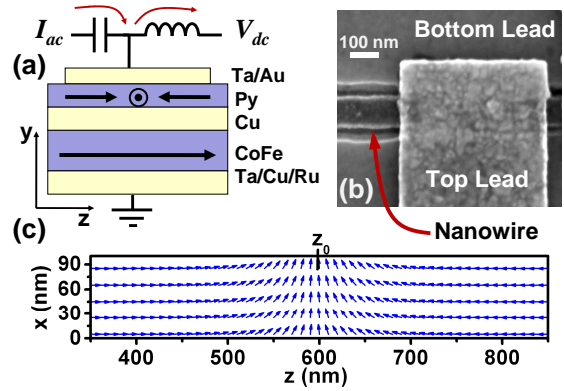}
\caption{(a) Schematic of a Co$_{50}$Fe$_{50}$/Cu/Py spin valve in the shape of a 90
nm wide nanowire with a 500 nm wide top lead.  A domain wall in the Py layer is
trapped at a random pinning site under the top lead while the Co$_{50}$Fe$_{50}$
layer is uniformly magnetized.  (b) Scanning electron microscopy image of the
device, showing the spin valve nanowire, the top lead and the bottom lead. (c)
Transverse domain wall profile in the Py nanowire obtained from micromagnetic
simulations.}
\end{figure}
Fig. 2(a) shows the resistance of the device as a function of magnetic field applied
parallel to the nanowire.  The giant magnetoresistance (GMR) data in Fig. 2(a,b)
reveal that the Py layer undergoes reversal in the +200 to -300 Oe field range
whereas the Co$_{50}$Fe$_{50}$ layer switches at -900 Oe.  Fig. 2(b) shows a major
GMR hysteresis loop of the Py layer ($\pm$600 Oe). An intermediate resistance (IR)
state in this loop arises from DW trapping near the middle of the top lead. The DW
is trapped in the same state in a minor hysteresis loop ($\pm$200 Oe) in Fig. 2(c).
All measurements reported in this Letter are made at $T$ = 80K in order to prevent
thermally-activated depinning of the DW from the IR state observed at 300K. We
expect the DW velocities at $T$ = 300K to be similar to those measured at $T$ = 80K
\cite{martinez}. For the thin-film nanowire geometry of the the Py layer, only a
transverse DW is expected to be stable \cite{beach}. We confirm this through
micromagnetic simulations using OOMMF code \cite{donahue}. The equilibrium
configuration of the DW in the Py layer given by the OOMMF simulations is shown in
Fig. 1(c).

We excite motion of the DW trapped in the IR state by applying an alternating
current with rms amplitude $I_{ac}$ between the top and the bottom leads of the
device. This current applies ac ST to the DW and induces oscillations of the DW
position along the nanowire. The DW oscillations give rise to spin valve resistance
oscillations, $\delta R_{ac}$, due to the GMR effect, and a rectified voltage,
$V_{dc}\sim I_{ac} \delta R_{ac}$, is generated by the device. Measurements of
$V_{dc}$ give information on the current-driven DW velocity. In our measurements, we
sweep the frequency of the ac current, $f$, in the 0.5 -- 10 GHz range and measure
$V_{dc}$ as a function of frequency \cite{tulap, sankey}.  The amplitude of the DW
oscillations reaches a maximum at the resonance frequency, $f_{0}$, of the DW
determined by the strength of the pinning potential, and a resonance peak is
observed in the measured response curve $V_{dc}(f)$.  

Figure 2(d) shows $V_{dc}(f)$ response curves measured in the middle of the field
range where the IR state is stable ($H$ = -50 Oe) with the DW trapped in the pinning
potential and at $H$ = 1 kOe with uniform magnetization of both Py and
Co$_{50}$Fe$_{50}$ layers. The -50 Oe response curve shows a resonance at 4 GHz due
to the DW oscillations in the pinning potential. We also observe a low-frequency
signal, which we attribute to the DW creep in inhomogeneities of the pinning
potential \cite{rebei, cayssol}.

The dependence of the DW velocity on current is given by the measurements of
$V_{dc}(f)$ as a function of $I_{ac}$. The maximum DW velocity $v_{max} = 2 \pi
f_{0} \Delta z(f_{0})$, where $\Delta z(f_{0})$ is the amplitude of DW oscillations
at resonance, is obtained from the data in Fig. 2(d) \cite{sankey}:
\begin{equation}
\label{amplitude}
\Delta z(f_{0}) = \frac{\sqrt{2}V_{dc}(f_{0})}{I_{ac}} \frac{L_c}{\Delta R}.
\end{equation}
where $L_{c}$=500 nm is the width of the top lead and $\Delta R$=37 m$\Omega$ is the
resistance difference between the parallel and antiparallel states of the spin
valve. 

\begin{figure}
\includegraphics[width=\columnwidth]{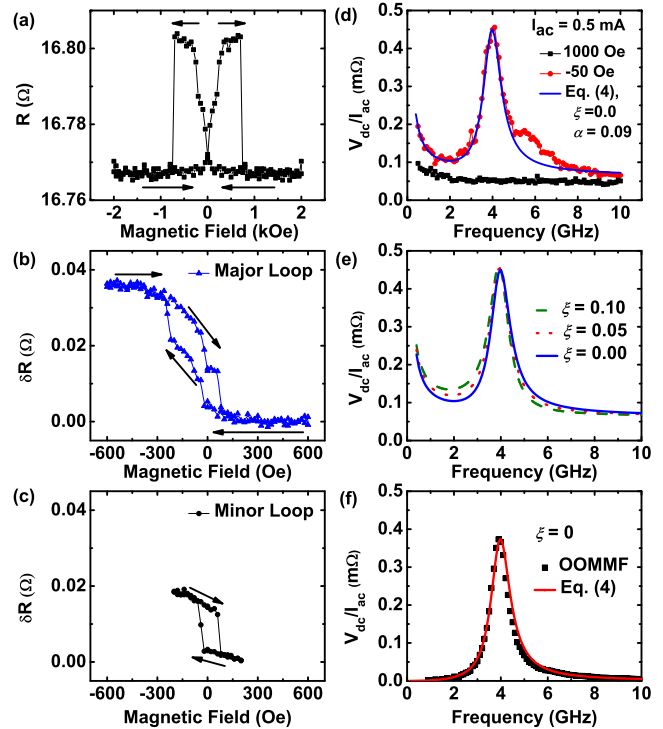}
\caption{(a) Magnetoresistance hysteresis loop of the Co$_{50}$Fe$_{50}$ /Cu/Py
nanowire spin valve for magnetic field applied parallel to the nanowire axis. The Py
layer undergoes magnetization reversal in the +200 to -300 Oe field range whereas
the Co$_{50}$Fe$_{50}$ layer switches abruptly at -900 Oe. Major (b) and minor (c)
hysteresis loops of the Py layer demonstrate that a DW (intermediate resistance
state) can be trapped near the middle of the top lead in a field range from -250 Oe
to +100 Oe. (d) $V_{dc}(f)$ response curve measured at -50 Oe with the DW trapped
under the top lead and at 1 kOe with uniformly magnetized Py layer.  A  resonance at
4 GHz is observed when the DW is present under the top lead. The solid line shows a
response curve fit calculated from Eq. (\ref{eqmotion}) with an added constant plus
1/f background. This fit assuming zero field-like torque ($\xi$ = 0) gives the best
fit parameters $P$=0.65, $\Lambda$=2.0 and $\alpha$=0.09. (e) Response curves
calculated with the same parameters as in (d) except for non-zero values of the
field-like torque ($\xi>$0). (f) The response curves calculated from
Eq.(\ref{eqmotion}) and from micromagnetic simulations at $I_{ac}$=0.5 mA are nearly
identical demonstrating the validity of the rigid DW approximation for
small-amplitude DW oscillations.}
\end{figure}

We compare our data to a theoretical description of DW dynamics in the rigid DW
approximation \cite{tatara1}. The motion of magnetization under the influence of ST
is described by the Landau-Lifshitz-Gilbert-Slonczewski equation \cite{thia2, khval,
slonc2}:

\begin{equation}
\label{LLGS}
\frac{d\vec{m}}{dt} = -\gamma \vec{m}\times \left(\vec{H}_{e} -
\frac{\alpha}{\gamma} \frac{d\vec{m}}{dt} + a_j \tau (\vec{m}\times \vec{p})\\ + 
b_j \vec{p}\right)
\end{equation}
where $\vec{m}$ is the unit vector in the direction of magnetization, $\vec{H}_{e}$
is the effective field, $\alpha$ is the Gilbert damping constant of Py,
\textit{a$_j$} = $\frac{\hbar J P}{2 d e M_s}$ is the ST coefficient,
$\tau=\frac{2\Lambda^2}{(\Lambda^2+1) +
(\Lambda^2-1)(\vec{m}\cdot \vec{p})}$ \cite{slonc2}, $\Lambda$ is the ST asymmetry
parameter, \textit{J} is the current density, \textit{P} is the spin polarization of
the current, \textit{d} is the Py layer thickness, \textit{e} is the electron
charge, $\gamma$ is the gyromagnetic ratio, $\vec{p}$ is the unit vector in the
direction of current polarization and \textit{b$_j$} = $\xi$\textit{a$_j$} is the
field-like torque coefficient. The dimensionless parameter $\xi$ describes the ratio
of the field-like torque to ST \cite{brataas}. The spatial profile of magnetization
in the DW is given by \cite{thia2}: 
\begin{subequations}
\label{dwshape}
\begin{eqnarray}
m_z = \cos\left(\theta(z)\right) = \tanh\left(\frac{z_0-z}{\lambda}\right)\\
m_x = \sin(\theta(z))\cos(\phi) =
\mbox{sech}\left(\frac{z_0-z}{\lambda}\right)\cos(\phi)\end{eqnarray}
\end{subequations}
Here we use a Cartesian system of coordinates with the $z$-axis parallel to the
nanowire, the $x$-axis in the plane of the Py film perpendicular to the nanowire,
and the $y$-axis along the Py film normal as shown in Fig. 1. In Eq.(\ref{dwshape}),
\textit{z$_{0}$} is the coordinate of the DW center, $\lambda$ is the DW width
($\lambda$= 53 nm is given by OOMMF simulations) and $\phi$ is the out-of-plane tilt
angle of magnetization at the DW center. In the rigid DW approximation ($\lambda =
const$), Eq. (\ref{LLGS}) can be rewritten in terms of the DW collective
coordinates, \textit{z$_{0}$} and $\phi$ \cite{tatara1, khval, boone3}:
\begin{subequations}
\label{eqmotion}
\begin{eqnarray}
\frac{\dot{z_0}}{\lambda}-\alpha\dot{\phi} - \gamma a_j\zeta_1(\Lambda) =
\frac{\gamma}{M_s}K \sin(2\phi)\\
-\alpha \frac{\dot{z_0}}{\lambda} - \dot{\phi} + \gamma b_j -\gamma
a_j\zeta_2(\Lambda)\tan(\phi)= \gamma  \frac{\partial H_p}{\partial z} z_0        
\end{eqnarray}
\end{subequations}
where \textit{K$ = 2 \pi M_{s}^{2} - K_{\bot} $} is the easy-plane shape anisotropy
constant, $\frac{\partial H_p}{\partial z}$ is the curvature of the pinning
potential, $K_{\bot}$ is the easy-axis perpendicular anisotropy constant
\cite{rant}, $\zeta_1(\Lambda)=\frac{2 \Lambda^2 ln[\Lambda]}{\Lambda^2-1}$ and
$\zeta_2(\Lambda)=\frac{\Lambda(\Lambda-1)}{\Lambda + 1}$ describe the effects of ST
asymmetry \cite{boone3}. The value of $K_{\bot}\approx0.4(2 \pi M_{s}^{2})$ is
determined from the measurements of resistance versus magnetic field applied
perpendicular to the sample plane. Assuming a parabolic pinning potential
$\left(\frac{\partial H_p}{\partial z}=const\right)$ and $\phi\ll1$,
Eq.(\ref{eqmotion}) can be solved analytically in the absence of spin torque and
damping. This solution describes free oscillations of the DW center, $z_0$ = $\Delta
z$ sin(2$\pi$$f_{0} t$), in the the pinning potential. The resonance frequency of
the DW, $f_0$, is determined by the pinning potential curvature
\cite{beach,tatara1,bedau}:
\begin{equation}
\label{potential}
\frac{\partial H_p}{\partial z} = \frac{2\pi^2 M_s f_0 ^2}{\lambda \gamma^2 K}. 
\end{equation}

The solid line in Fig. 2(d) shows a fit of a theoretical response curve $V_{dc}(f)$
to the experimental data. The theoretical response curve is found by numerically
solving Eq.(\ref{eqmotion}) and adding a constant plus $1/f$ noise background. The
fitting parameters used in the fitting procedure are $\alpha$, $P$ and $\Lambda$.
The fit assuming zero field-like torque ($\xi=0$) gives $P$=0.65, $\Lambda$=2.0 and
$\alpha$=0.09. This value of the damping parameter significantly exceeds the typical
Py value $\alpha\approx$0.01. We directly measure damping of the lowest-energy spin
wave mode of the Py nanowire in the state of saturated magnetization using a ST
ferromagnetic resonance technique \cite{tulap, sankey, boone} and find
$\alpha$=0.09. This shows that the high value of $\alpha$ is an intrinsic property
of the ultra-thin Py film and does not originate from emission of spin waves by the
moving DW. The observed large $\alpha$ can be explained by diffusion of Cu into the
Py film known to increase damping \cite{guan} and by spin pumping important for
ultra-thin ferromagnetic films \cite{heinrich}. The theoretical fit curve in Fig.
2(e) shows deviations from the experimental data at frequencies above the DW
resonance. We attribute these deviations to excitation of spin waves in the Py
nanowire. We calculate the bottom of the dispersion relation of the lowest-frequency
spin wave mode in the Py nanowire to be at 5.5 GHz \cite{boone} in agreement with
the data in Fig. 2(d).

Figure 2(e) shows theoretical response curves for a non-zero field-like term
($\xi>0$) in the equations of motion. Our calculations show that a non-zero
field-like torque adds an antisymmetric Lorentzian component to the response curve
and decreases the quality of the theoretical fit to the experimental data. Fig. 2(e)
illustrates that theoretical $V_{dc}(f)$ curves with $\xi>0.1$ show substantial
deviations from our experimental data and thus our measurements set an upper limit
on the value of field-like torque in our spin valve structure.  

We also compare our experimental data to results of micromagnetic simulations. The
squares in Fig. 2(f) give the response curve obtained from OOMMF simulations. In
these simulations, a spatially-nonuniform field $H_{p} = -\frac{\partial
H_{p}}{\partial z} z$ that mimics the 4 GHz pinning potential is applied parallel to
the Py nanowire and ST from $I_{ac}$ = 0.5 mA is applied to drive oscillations of
the DW in the nanowire. The solid line in Fig. 2(f) obtained by numerically solving
Eq.(\ref{eqmotion}) is in a good agreement with the micromagnetic simulation
results.

\begin{figure}
\includegraphics[width=\columnwidth]{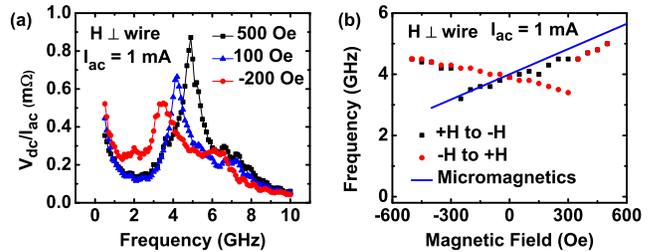}
\caption{(a) $V_{dc}(f)$ response curves measured at several magnetic fields,
$H_{\bot}$, applied in the plane of the Py layer perpendicular to the nanowire axis.
(b) Frequency of the DW resonance as a function of $H_{\bot}$. Solid line is
obtained from micromagnetic simulations. The hysteresis in the DW resonance
frequency as a function $H_{\bot}$ is due to the field-induced reversal of the DW
chirality at $H_{\bot}$$\approx$350 Oe.}
\end{figure}

To further confirm that the observed resonance in $V_{dc}(f)$ is due to DW
oscillations and not due to spin wave excitations, we measure the resonance
frequency as a function of magnetic field, $H_{\bot}$, applied in the plane of the
Py film perpendicular to the nanowire axis. This magnetic field modifies the
curvature of the pinning potential, thereby changing the DW resonance frequency
\cite{saitoh}.  Fig. 3(a) shows $V_{dc}(f)$ measured at several values of
$H_{\bot}$, and Fig. 3(b) shows the dependence of $f_0$ on $H_{\bot}$. The resonance
frequency shifts with $H_{\bot}$, and the frequency shift is odd in $H_{\bot}$. This
odd frequency shift is a clear signature of the DW resonance for which the sign of
the shift is determined by the DW chirality \cite{saitoh}. A frequency shift
symmetric with respect to $H_{\bot}$ is expected from spin wave resonances in this
geometry and thus the data in Fig.(3) exclude the possibility of spin wave origin of
the observed resonance. The hysteretic jumps of $f_0$ at $H_{\bot}\approx\pm 350$ Oe
correspond to switching of the DW chirality induced by $H_{\bot}$. The dependence of
$f_0$ on $H_{\bot}$ given by micromagnetic simulations is shown in Fig. 3(b). It is
in good agreement with the experimental data. 

The dependence of the DW velocity on current is given by measurements of $V_{dc}(f)$
as a function of $I_{ac}$ shown in Fig. 4(a). Fig. 4(b) shows the maximum
instantaneous DW velocity, $v_{max}$, versus current density, $J_{ac}$, calculated
from the data in Fig. 4(a) using Eq.(\ref{amplitude}).  We observe velocities as
high as 800 m/s for $J_{ac}$=9$\times$10$^6$ A/cm$^2$.  The amplitude of the DW
oscillations at $v_{max}$ = 800 m/s is 32 nm while the smallest DW oscillation
amplitude detectable with our technique is 1 nm.

Because the DW mass $\mu \sim M_s$ and the ST coefficient $a_j \sim 1/M_s$, we
expect the DW velocity to scale as $1/M_s^2$. The DW velocity at resonance is also
expected to be inversely proportional to damping $v_{max}\sim 1/\alpha$. Therefore,
for standard Py (Ni$_{80}$Fe$_{20}$) material parameters ($M_s$=800 emu/cm$^3$,
$\alpha = 0.01$), the expected maximum oscillating DW velocity in the CPP geometry
is approximately 2000 m/s at the current density of 10$^7$ A/cm$^2$. DW velocities
as high as 325 m/s at a current density of $5\times10^7$ A/cm$^2$ were observed in
the current-in-plane geometry for samples with the standard Py material parameters
\cite{bocklage}. 

\begin{figure}
\includegraphics[width=\columnwidth]{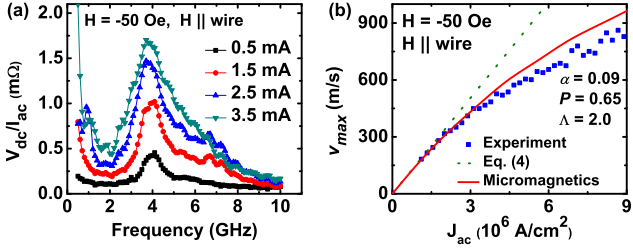}
\caption{(a) $V_{dc}(f)$ response curves  measured at several values of the
excitation current, $I_{ac}$. (b) Maximum speed of the oscillating DW in the pinning
potential versus the rms ac drive current density, $J_{ac}$.  Squares are the
experimental data, dashed line is the rigid DW approximation result given by Eq.
(\ref{eqmotion}), solid line is given by micromagnetic simulations.}
\end{figure}

The rigid DW approximation predicts a linear increase of the DW velocity with
$J_{ac}$ but the experimental data in Fig. 4(b) show clear deviations from the
linear behavior for $J_{ac}>$1.5$\times$10$^6$ A/cm$^2$.  The origin of these
deviations is revealed by micromagnetic simulations. The results of these
simulations for $v_{max}(J_{ac})$ are shown in Fig. 4(b). The sublinear dependence
of the DW velocity on current similar to that observed in the experiment arises from
the breakdown of the rigid DW approximation. The micromagnetic simulations show
significant deformations of the DW shape induced by the pinning potential at large
amplitudes of DW oscillations.

In conclusion, we make measurements of DW dynamics in Py nanowires excited by
spin-polarized current applied perpendicular to the nanowire. In this geometry, we
observe DW velocities as high as 800 m/s at current densities of 9$\times$10$^6$
A/cm$^2$.  The high DW velocities excited by current applied perpendicular to
ferromagnetic nanowires are promising for nonvolatile memory and microwave signal
processing applications \cite{braga}. This work was supported by the NSF Grants
DMR-0748810 and ECCS-0701458, and by the Nanoelectronics Research Initiative through
the Western Institute of Nanoelectronics.



\end{document}